\documentstyle[prd,aps,floats]{revtex}

\begin{document}
\preprint{SUSSEX-AST 97/5-1, FERMILAB-Pub-97/139-A, astro-ph/9705166}
\draft

%
%
\input epsf
\renewcommand{\topfraction}{0.8}
\twocolumn[\hsize\textwidth\columnwidth\hsize\csname 
@twocolumnfalse\endcsname

\title{Primordial black hole constraints in cosmologies with early matter 
domination} 
\author{Anne M.~Green and Andrew R.~Liddle}
\address{Astronomy Centre, University of Sussex, Falmer, Brighton BN1
9QH,~~~U.~K.}  
\author{Antonio Riotto}
\address{NASA/Fermilab Astrophysics Center, Fermi National Accelerator 
Laboratory, Batavia, IL~~60510}
\date{\today} 
\maketitle
\begin{abstract}
Moduli fields, a natural prediction of any supergravity and 
superstring-inspired supersymmetry theory, may lead to a  prolonged period 
of matter domination in the early Universe. This can be observationally 
viable provided the moduli decay early enough to avoid harming 
nucleosynthesis. If primordial black holes form, they would be expected to 
do so before or during this matter dominated era. We examine the extent to 
which the standard primordial black hole constraints are weakened in such a 
cosmology. Permitted mass fractions of black holes at formation are of order 
$10^{-8}$, rather than the usual $10^{-20}$ or so. If the black holes form 
from density perturbations with a power-law spectrum, its spectral index is 
limited to $n \lesssim 1.3$, rather than the $n \lesssim 1.25$ obtained in 
the standard cosmology.
\end{abstract}

\pacs{PACS numbers: 98.80.Cq \hspace*{0.3cm} SUSSEX-AST 97/5-1, 
FERMILAB-Pub-97/139-A, astro-ph/9705166}

\vskip2pc]

\section{Introduction}

Although a substantial amount of work has been carried out on the assumption 
of a `standard' cosmology, in which the Universe proceeds from an early 
period of inflation through reheating to radiation domination and finally to 
matter domination in the recent past, there is no direct evidence supporting 
this picture until the relatively late epoch at which nucleosynthesis 
occurs. Recently, this standard picture has been questioned and some 
alternative cosmologies discussed. An example is `thermal inflation' 
\cite{LS:TI,moreTI}, a short second period of inflation at lower energy 
scales which does not generate interesting density perturbations, but which 
may resolve additional relic density problems not solved by the original 
inflationary period.

Several cosmological constraints are sensitive to whatever assumption is 
made for the entire cosmological evolution; axion cosmology is one such 
situation \cite{LS:TI}, and the constraints originating from primordial 
black holes (PBHs) is another. Recently, the latter was reinvestigated for 
cosmologies with thermal inflation, showing that the standard constraints 
\cite{var:lim,Carr:rev} on the formation density of PBHs would weaken quite 
markedly \cite{GL}.

Another possible modification to the standard cosmology is the addition of a 
prolonged period of matter domination, induced by a slow-decaying massive 
particle. In  $N=1$ supergravity models \cite{sugra}, supersymmetry (SUSY) 
is broken in some hidden sector and the gravitational strength force plays 
the role of messenger by transmitting SUSY breaking down to the visible 
sector. In these models there often exist scalar fields with masses of the 
order of the weak scale and gravitational strength coupling to the ordinary 
matter. If at early epochs one of these fields is sitting far from the 
minimum of its potential with an amplitude of order of the Planck scale, 
the coherent oscillations about the minimum will eventually dominate the 
energy density of the universe. These fields will then behave like 
nonrelativistic matter, and decay at very late times. The presence of these 
slow-decaying massive particles is predicted  not only in some specific 
classes of  supergravity models, but in almost all theories in which 
supersymmetry is broken at an intermediate scale. In string models, massless 
fields exist in all known string ground states and parametrize the 
continuous ground state 
degeneracies characteristic of supersymmetric theories. These fields are 
massless to all orders in perturbation theory, and get their mass, of order 
a TeV, from the same non-perturbative mechanism which breaks SUSY.  Being 
coupled to the ordinary matter only by gravitational strength couplings, a 
long lifetime results. Possible examples are the dilaton of string theory 
and the massless gauge singlets of string compactifications, and they go 
generically under the name of moduli. Under natural assumptions on the 
couplings, one finds that the reheating temperature after the moduli decay 
is too low to allow standard nucleosynthesis. Therefore, moduli are 
generally far {\em too} good at giving a period of matter domination, 
lasting beyond the epoch of nucleosynthesis and destroying this crucial 
success of the standard cosmology \cite{moduli1,moduli2}.

Many attempts have been made to resolve this cosmological moduli problem 
\cite{attempts}. However, all of them require new phenomena to occur on the 
cosmological side as well as in the theory of supersymmetry breaking. It is 
not our intention, in this paper, to propose another solution to the moduli 
problem, but rather to study the extent to which the modular cosmology may 
affect the standard primordial black hole constraints. We will therefore 
assume that  the moduli are somewhat more massive than the supersymmetry 
scale set by the gravitino mass $m_{3/2}=(10^2-10^3) \,$GeV, and that their 
decay can be just early enough. This assumption seems reasonable in view of 
the recent developments in the context of the field-theory limit of 
superstrings \cite{gb}, where it has been shown that moduli masses as large 
as $10^3 \, m_{3/2}$ can be achieved without incurring excessive 
unnaturalness. The cosmological sequence in such a model is as 
follows. At a high temperature the moduli come to dominate and the Universe 
begins an epoch of matter--domination. During this period, the radiation 
field actually cools to some way below the nucleosynthesis scale (about 
$10^{-3}\,$GeV), but the moduli decay while the energy density is still high 
enough to permit thermalization slightly above the nucleosynthesis 
temperature. In this picture, baryogenesis must be caused by the decay of 
the moduli, rather than at the electro-weak transition \cite{b}. 

\section{The moduli--dominated epoch}

In hidden-sector models, supersymmetry breaking is conveyed to the 
low-energy visible sector through Planck scale suppressed interactions. In 
non-renormalizable hidden-sector models, supersymmetry vanishes in the limit 
$m_{{\rm Pl}} \rightarrow \infty$, $m_{{\rm Pl}}$ being the Planck mass. 
Since the potential for a generic moduli field $\phi$  is generated through 
the same physics associated to supersymmetry breaking, its potential takes 
the form
\begin{equation}
V(\phi)=m_{3/2}^2 \: M_{{\rm Pl}}^2 \: {\cal V}(|\phi|/M_{{\rm Pl}}),
\end{equation}
where $M_{{\rm Pl}}=m_{{\rm Pl}}/\sqrt{8\pi}$ is the reduced Planck mass and 
$m_{3/2}\sim 1\,$TeV is the gravitino mass. The potential for this dangerous 
direction vanishes in the flat-space limit since $m_{3/2}\rightarrow 0$ in 
that limit. As mentioned in the introduction, excitations around the 
zero-temperature minimum $\phi_0$ of the potential have a mass $m_\phi={\cal 
O}(10^3) \, m_{3/2}$. 

Moduli fields are expected to be initially shifted from their 
zero-temperature minimum due to the effect of thermal fluctuations or of 
quantum fluctuations during inflation \cite{dis}. Another source of the 
shift might be the fact that the moduli couplings to the inflaton generally 
modify, during inflation, the properties of the effective potential. Moduli 
usually acquire a mass squared of the order of $H^2$, where $H\sim 
10^{13}\,$GeV is the Hubble parameter during the inflationary stage, and the 
value of the minimum of the potential  may be shifted \cite{sh}. The shift 
produced by such effects may be as large as $m_{{\rm Pl}}$.

Although the form of the potential is not known, for our purposes one may 
just consider  
oscillations around the minimum with initial amplitude $\phi_{{\rm i}}$ and 
take $V(\phi)\simeq m_\phi^2\phi^2/2$. When the Hubble parameter $H$ reaches 
a value $H\sim m_\phi$, the scalar field starts oscillating coherently 
around the minimum of the potential. This happens when the temperature of 
the universe is $T_{{\rm i}}\sim \sqrt{m_\phi m_{{\rm Pl}}}$ (in the case in 
which the universe is radiation dominated at that epoch).

The initial energy stored in the oscillations $\rho_{{\rm i}} \sim m_\phi^2 
\phi_{{\rm i}}^2$ redshifts like matter and can eventually dominate the 
energy density. When it does so depends on $\phi_{{\rm i}}$. If $\phi_{{\rm 
i}}$ is of order $m_{{\rm Pl}}$, then the moduli dominate immediately, while 
if $\phi_{{\rm i}}$ is smaller, radiation domination will continue for a 
while before the moduli come to dominate, or, in extreme cases, the moduli 
may decay before they dominate the energy density. In hidden-sector models, 
moduli couple to other fields only through Planck suppressed interactions. 
Examples of such fields are the dilaton and the compactification moduli of 
string theory or, in general, for any gauge singlet field responsible for 
SUSY breaking. There are several types of Planck suppressed couplings the 
moduli might have with ordinary matter, but all of them lead to the same 
estimate of the decay width
\begin{equation}
\Gamma_\phi\sim \frac{m_\phi^3}{m_{{\rm Pl}}^2}.
\end{equation}
The condition for the moduli to dominate the energy density of Universe when 
they decay (which will be at the epoch $H \simeq \Gamma_\phi$) is that their 
initial value satisfies
\begin{equation}
\phi_{{\rm i}} \gtrsim 10^{-8} \, \left( \frac{m_\phi}{1\,{\rm TeV}}
	\right)^{1/2} \, m_{{\rm Pl}} \,.
\end{equation}
At the decay time the radiation fluid has temperature
\begin{equation}
T_{{\rm dec}} \sim m_\phi^{11/6} \phi_{{\rm i}}^{-1/6}  
	m_{{\rm Pl}}^{-2/3} \,.
\end{equation} 
The decay products of the moduli will thermalize, reheating the universe 
up to a temperature 
\begin{equation}
T_{{\rm reh}} \sim \frac{m_\phi^{3/2}}{m_{{\rm Pl}}^{1/2}} \sim 
	3 \times 10^{-4} \left(\frac{m_\phi}{10^2 \, {\rm GeV}}
	\right)^{3/2} \, {\rm MeV}  \,.
\end{equation} 
Notice that the reheating temperature is independent of $\phi_{{\rm i}}$, 
provided that the universe is dominated by the moduli energy density when 
decays start. 

The decay products of $\phi$ will destroy the $^4$He and D nuclei, and thus 
successful nucleosynthesis predictions, unless $T_{{\rm reh}}$ is larger 
than about 1~MeV. If the moduli field has mass $10^2\,$GeV, $T_{{\rm reh}}$ 
is well below the energy scale of nucleosynthesis, but if instead one 
assumes $m_\phi \sim 10^4\,$GeV, then the reheat temperature becomes 
comparable and it may be possible to thermalize to a high enough temperature 
for standard nucleosynthesis to proceed. In the case $\phi_{{\rm i}} 
\sim m_{{\rm Pl}}$ where the moduli dominate as soon as they being to 
oscillate, this corresponds to an expansion of the Universe during matter 
domination by a factor of around $(m_{{\rm Pl}}/m_\phi)^{4/3} \sim 10^{20}$, 
a very prolonged period indeed.

\section{Primordial black hole constraints}

\subsection{Formation density constraints}

In a radiation--dominated Universe at temperature $T$, the horizon mass is 
given roughly by
\begin{equation}
\label{mhor}
M_{{\rm H}} \simeq 10^{18} \, {\rm g} \, \left( \frac{10^7 \, {\rm
	GeV}}{T} \right)^2 \,.
\end{equation}
PBHs of a given mass are expected to form around the time when that mass 
equals the horizon mass; production of smaller black holes is suppressed as 
pressure prevents the collapse of any density perturbation. In a 
matter-dominated Universe, formation may occur on scales below the horizon 
mass, as we discuss later.

The lifetime of the black hole can be parametrized as 
\cite{Page:tevap,Carr:rev}
\begin{equation}
\label{tau}
\tau_{\rm{evap}} = \frac{9 \times 10^{-27}}{f(M)} \left( 
	\frac{M}{1\,\rm{g}}\right)^3 {\rm sec} \,,
\end{equation}
where $f(M)$ depends on the number of particle species which can be
emitted and is normalized to 1 for holes which emit only massless
particles. A black hole of initial mass around $5 \times 10^{14}\,$g would 
be evaporating at the present epoch, while masses around $10^{10}\,$g would 
be evaporating at nucleosynthesis. Those lighter black holes may form early 
on in the period of moduli domination, or even before it if its onset is 
delayed.

We denote the fraction of the density of the Universe in black holes of a 
given mass as $\beta$, with $\beta_{{\rm i}}$ denoting the initial density 
at formation. The ratio of the PBH density to the density in other forms is 
denoted $\alpha \equiv \beta/(1-\beta)$.

The various limits which can be placed on the PBH density are well known 
\cite{var:lim,Carr:rev}; we shall use the compilation given in 
Ref.~\cite{GL}. There are a range of constraints from effects of 
evaporation, while for more massive black holes, $M \gtrsim 10^{15}\,$g, the 
only limit comes from their contribution to the present density parameter. 
An additional, less secure constraint arises if one assumes that evaporation 
leaves behind a Planck mass relic \cite{MacG}.

\begin{figure}[t]
\centering 
\leavevmode\epsfysize=6cm \epsfbox{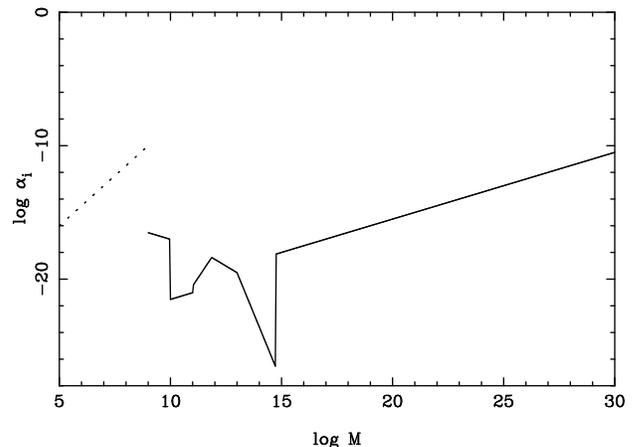}\\ 
\caption[fig1]{\label{fig1}  The tightest limits on $\alpha_{{\rm i}}$, for 
the standard cosmology. The mass is in grams. The relic constraint is shown 
as a dotted line, emphasizing that it is not compulsory. The solid lines, 
from left to right, represent ${\rm n\bar{n}}$ production at 
nucleosynthesis, deuterium destruction, He-4 spallation, entropy production, 
gamma ray production, and total density in PBHs; see Ref.~\cite{GL} for 
details.}
\end{figure}

All these constraints are expressed as limits on the fraction of the mass of 
the Universe in black holes at the present or at the time of evaporation. To 
constrain the initial mass fraction, one needs to assume a form for the 
entire cosmology back to the formation epoch, given by Eq.~(\ref{mhor}). 
Fig.~\ref{fig1} shows the result of carrying this out for the standard 
cosmology, where the Universe was radiation dominated until very recently 
\cite{GL}. We see that the constraints are extremely tight; 
typically only something like $10^{-20}$ of the mass of the Universe is 
permitted to form black holes in the standard cosmology.

We now turn to our main purpose, examining the change in the constraints 
induced by a period of moduli domination. In order to attain a reheating 
temperature of $T_{{\rm reh}} \sim 10^{-3}\,$GeV, so that nucleosynthesis 
can proceed, we require $m_{\phi} \sim 2 \times 10^{4}\,$GeV. In the extreme 
case of $\phi_{{\rm i}} \sim m_{{\rm Pl}}$, the moduli begin to 
oscillate at temperature $T_{{\rm MD}} \sim 2 \times 10^{11}\,$GeV, since
\begin{equation}
\rho_{{\rm i}} = m_{\phi}^2 m_{{\rm Pl}}^2 = \frac{\pi^2}{30}
	g_{\star}^{{\rm MD}} T_{{\rm MD}}^4 \,.
\end{equation}
Here $g_{\star}^{{\rm MD}} \sim 250$ is the number of degrees of freedom in 
the minimal supersymmetric standard model.  In the following, we will 
consider two scenarios: one in which $\phi_{{\rm i}} \sim m_{{\rm Pl}}$ and 
moduli domination begins immediately, and an intermediate scenario where 
$\phi_{{\rm i}} \sim 10^{-4} \, m_{{\rm Pl}}$ and there is a delay before 
moduli domination commences.

\subsubsection{Immediate moduli domination}

{}From Eq.~(\ref{mhor}), PBHs in the mass range $ 2 \times 10^{9} \, {\rm g} 
\leq M \leq 2 \times 10^{38} \, {\rm g}$ are formed
during the moduli-dominated era. During moduli-domination, the PBHs 
constitute a constant fraction of the total energy density. The time of the 
decay of the moduli, $t_{{\rm dec}}$, and of the reheating of the subsequent 
thermalized fluid, $t_{{\rm reh}}$, can be taken as the same, giving a PBH 
density at reheating of
\begin{equation}
\left( \frac{\rho_{{\rm PBH}}}{\rho_{{\rm rad}}} \right)_{{\rm reh}} 
	= \left( \frac{\rho_{{\rm PBH}}}{\rho_{{\rm mod}}}
	\right)_{{\rm dec}} = \left( \frac{\rho_{{\rm PBH}}}
	{\rho_{{\rm mod}}} \right)_{{\rm i}} \equiv \alpha_{{\rm i}} \,.
\end{equation}
Here $\rho_{{\rm rad}}$ and $\rho_{{\rm mod}}$ are the energy densities 
in radiation and moduli respectively. Therefore, for PBHs formed during 
moduli domination and surviving beyond it (which is all those of interest), 
we have
\begin{equation}
\left( \frac{\rho_{{\rm PBH}}}{\rho_{{\rm rad}}} \right)_{{\rm evap}} 
	\equiv \alpha_{{\rm evap}} = \frac{\beta_{{\rm i}}}{ 
	1-\beta_{{\rm i}}} \frac{T_{{\rm reh}}}{T_{{\rm evap}}} \,.
\end{equation}
Considering the duration of the various phases of the evolution
of the universe
\begin{equation}
\frac{t_{{\rm evap}}}{t_{{\rm Pl}}} = \frac{t_{{\rm evap}}}{t_{{\rm reh}}}
	\frac{t_{{\rm dec}}}{t_{{\rm i}}} \frac{ t_{{\rm i}}}
	{t_{{\rm Pl}}} \,,
\end{equation}
using the relation between the formation time and mass of a PBH for PBHs
formed during radiation domination~\cite{Carr:rev}\footnote{This equation is 
not precisely valid for PBHs which are formed during matter domination since 
their formation may be delayed; however, it can be used in this context 
since the delay is negligible compared with the PBH lifetime. We will 
discuss PBH formation during matter domination in more detail later.}
\begin{equation}
M \cong M_{{\rm H}} = \frac{t_{{\rm i}}}{t_{{\rm Pl}}} M_{{\rm Pl}} \,,
\end{equation}
and the variation of the density during moduli domination, $\rho 
\propto t^{-2}$, leads to
\begin{equation}
\frac{t_{{\rm evap}}}{t_{{\rm Pl}}} = \left(\frac{T_{{\rm reh}}}
	{T_{{\rm evap}}} \right)^2  \left( \frac{ \rho_{{\rm i}}}
	{\rho_{{\rm dec}}} \right)^{1/2} \frac{M}{m_{{\rm Pl}}} \,.
\end{equation} 
In order to eliminate $\rho_{{\rm i}}$, Eq.~(\ref{mhor}) can be rewritten as 
\begin{equation}
M_{{\rm H}} = 0.2 \, \frac{m_{{\rm Pl}}^3}{\rho^{1/2}} \,.
\end{equation}
Finally the resulting expression for $t_{{\rm evap}}$ can be equated with 
Eq.~(\ref{tau}) to give
\begin{equation}
\frac{T_{{\rm reh}}}{T_{{\rm evap}}} = 8 \times 10^{-21} 
	\left(\frac{M}{m_{{\rm Pl}}} \right)^{3/2} \,,
\end{equation}    
so that 
\begin{equation}
\frac{\beta_{{\rm i}}}{1-\beta_{{\rm i}}} = 1 \times 10^{20} \left(
	\frac{m_{{\rm Pl}}}{M} \right)^{3/2} \alpha_{{\rm evap}} \,.
\end{equation}
The gravitational constraints, which require that the present-day densities 
of PBHs and relics not to overclose the universe, are
\begin{eqnarray}
\Omega_{{\rm PBH,eq}} &  = & \left( \frac{\rho_{{\rm PBH}}}{\rho_{{\rm 
	rad}}} \right)_{{\rm eq}} = \frac{\beta_{{\rm i}}}{ 1-
	\beta_{{\rm i}}} \frac{T_{{\rm reh}}}{T_{{\rm eq}}} <1 \\
\Omega_{{\rm rel,eq}} &  = & \left( \frac{\rho_{{\rm PBH}}}{\rho_{{\rm 
	rad}}} \right)_{{\rm eq}} =\frac{m_{{\rm Pl}}}{M}
	\frac{\beta_{{\rm i}}}{ 1-\beta_{{\rm i}}}
	\frac{T_{{\rm reh}}}{T_{{\rm eq}}} <1 \,,
\end{eqnarray}
where `eq' indicates the epoch of matter--radiation equality in the 
Universe's recent past, after which the density of PBHs or relics, relative 
to the critical density, remains constant.
In the case of PBHs with $M > 2 \times 10^{38}\,$g, formed after moduli
domination, the requirement that the present-day density of PBHs does
not overclose the universe is obviously the same as in the standard 
evolution of the universe. The PBHs formed before moduli domination are 
sufficiently light $M \leq  10^{9}\,$g that only the relic constraint 
$\Omega_{{\rm rel,eq}} < 1$ applies to them, where:
\begin{equation}
\Omega_{{\rm rel,eq}} =\left( \frac{\rho_{{\rm PBH}}}{\rho_{{\rm rad}}}
	\right)_{{\rm eq}} =\frac{m_{{\rm Pl}}}{M} 
	\frac{\beta_{{\rm i}}}{ 1-\beta_{{\rm i}}} 
	\frac{T_{{\rm reh}}}{T_{{\rm eq}}} \frac{T_{{\rm i}}}{T_{{\rm MD}}} 
	<1  \,,
\end{equation}
and in fact it turns out that large initial mass fractions of PBHs, 
$\beta_{{\rm i}} \sim 1$, are allowed.
 
\begin{figure}[t]
\centering 
\leavevmode\epsfysize=6cm \epsfbox{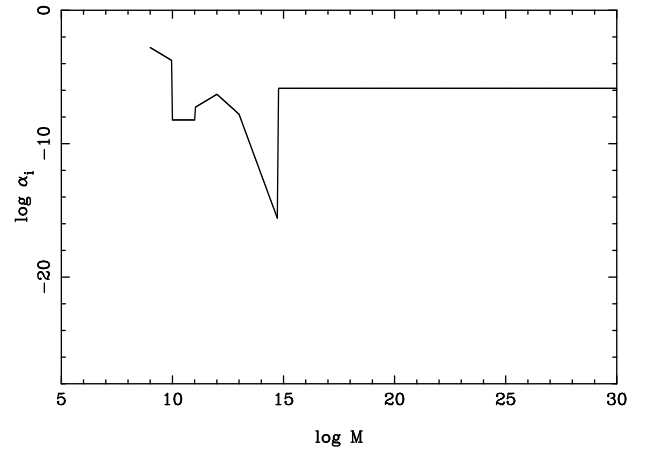}\\ 
\caption[fig2]{\label{fig2}  The tightest limits on the initial
mass fraction of PBHs, $\alpha_{{\rm i}}$, if moduli domination commences
immediately. The mass is in grams. The rightmost line, indicating the 
density constraint, continues horizontally until $M \sim 10^{38}\,$g; PBHs 
more massive than this form after moduli domination and the standard 
constraint $\alpha_{{\rm i}} < 10^{-19} \sqrt{M/10^{15} \, {\rm g}}$ then 
applies. The constraints are the same as in Fig.~1.}
\end{figure}

The various limits on the initial mass fraction of PBHs are illustrated in 
Fig.~\ref{fig2}.

\subsubsection{Delayed moduli domination}

An initial value $\phi_{{\rm i}} \sim m_{{\rm Pl}}$ is the most natural, but 
it is not impossible for it to be smaller and this leads to a shorter period 
of moduli domination. As an example, we take $\phi_{\rm i} \sim 10^{-4} 
m_{\phi}$ so that moduli domination commences when the energy stored in the 
oscillations of the moduli field becomes greater than that of the radiation,
\begin{equation}
\frac{\rho_\phi}{\rho_{{\rm rad}}} = \frac{\phi_{\rm i}^2}{m_{\rm 
Pl}^2}
	\frac{ 2 \times 10^{11} {\rm GeV}}{T_{\rm MD}} >1 \,,
\end{equation}
at temperature $T_{\rm MD} = 2\times 10^{3}$ GeV. 

{}From Eq.~(\ref{mhor}), PBHs with $M<2 \times 10^{25}\,$g  are formed in 
the radiation-dominated period before the moduli domination commences. Their
energy density, relative to that in other forms, varies as $T^{-1}$
initially then remains constant during moduli domination. It then increases
as $T^{-1}$ during the subsequent radiation domination so that
\begin{equation}
\left( \frac{\rho_{\rm PBH}}{\rho_{\rm rad}} \right)_{\rm evap} 
	\equiv \alpha_{\rm evap} = \frac{\beta_{\rm i}}
	{1 -\beta_{\rm i}} \frac{T_{\rm i}}{T_{\rm MD}} 
	\frac{T_{\rm reh}} {T_{\rm evap}} \,,
\end{equation}
leading to 
\begin{equation}
\frac{\beta_{\rm i}}{1 -\beta_{\rm i}} = 2 \times 10^{5} 
	\frac{m_{\rm Pl}}{M} \alpha_{\rm evap} \,.
\end{equation}
Similarly for the gravitational constraints
\begin{eqnarray}
\Omega_{\rm PBH,eq} & = & \left( \frac{\rho_{\rm PBH}}{\rho_{\rm rad}}
	\right)_{\rm eq} = \frac{\beta_{\rm i}}{1- \beta_{\rm i}}
	\frac{T_{\rm i}}{T_{\rm MD}} \frac{T_{\rm reh}}{T_{\rm eq}} <1 \\
\Omega_{\rm rel,eq} & = & \left( \frac{\rho_{\rm rel}}{\rho_{\rm rad}}
	\right)_{\rm eq} = \frac{m_{\rm Pl}}{M} \frac{\beta_{\rm i}}
	{1- \beta_{\rm i}} \frac{T_{\rm i}}{T_{\rm MD}}
	\frac{T_{\rm reh}}{T_{\rm eq}} < 1 \,.
\end{eqnarray}

For PBHs with $M> 2 \times 10^{25}\,$g, formed after moduli domination
commences, the gravitational constraint is the same as when moduli 
domination starts immediately at $T =  2\times 10^{11}\,$GeV.
 
\begin{figure}[t]
\centering 
\leavevmode\epsfysize=6cm \epsfbox{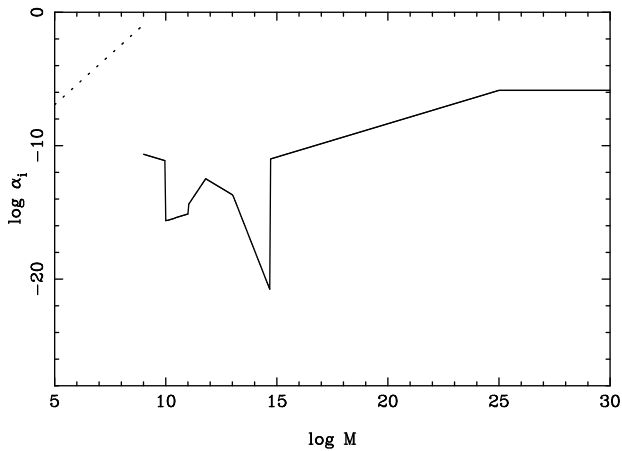}\\ 
\caption[fig1]{\label{fig3}  The tightest limits on the initial
mass fraction of PBHs, $\alpha_{\rm_{i}}$, if $\phi_{\rm i} \sim 10^{-4} 
m_{\rm Pl}$ and moduli domination is delayed. 
The mass is in grams. For $M > 2\times 10^{25}$g the limits are the
same as when moduli domination commences immediately. The constraints are
the same as in Fig.~1.}
\end{figure}

The various limits on the initial mass fraction of PBHs in this case are 
illustrated in Fig.~\ref{fig3}.

\subsection{Density perturbation constraints}

PBH formation can be used to constrain the spectral index of the density 
perturbation spectrum \cite{CGL:PBH,kim,GL}. To do this, we assume an 
initial 
spectrum which is a power-law across the entire range of scales from the PBH 
scale up to the present horizon scale. Constraints can then be placed on the 
spectral index $n$ of those perturbations. To do this,
we need to consider the variation of $\sigma_{{\rm hor}}(M)$, the mass 
variance evaluated at horizon crossing \cite{GL}. During matter domination 
$\sigma_{{\rm hor}}(M) \propto M^{(1-n)/6}$, whereas during radiation 
domination $\sigma_{{\rm hor}}(M) \propto M^{(1-n)/4}$; the different 
scalings arise because during matter-domination the comoving mass density is 
conserved but during radiation-domination it decreases so that 
mass scales enter the horizon more quickly \cite{GL}.

In the case of PBHs formed during matter 
domination the standard scenario
for PBH formation no longer holds. It has been shown \cite{cohere} that 
perturbation growth during coherent scalar field oscillation behaves in 
exactly the same way as in a dust Universe, provided, as here, that the 
oscillations are very rapid compared to other timescales in the problem. PBH 
formation during such a matter-dominated period 
was considered in Ref.~\cite{CGL:PBH}. Because there is no pressure, it is 
now possible for PBHs to form well within the horizon, but in order to do so 
the initial perturbation must be sufficiently spherical as gravitational 
collapse is unstable to aspherical growth.  The formation rate 
is given by \cite{YP:dust}
\begin{equation}
\beta(M) \approx 2 \times 10^{-2} \sigma^{13/2}(M) \,.
\end{equation}

For PBHs with $M> 2 \times 10^{9}\,$g, formed during moduli domination, we 
have
\begin{eqnarray}
\sigma_{{\rm hor}}(M) & =  & \sigma_{{\rm hor}}(M_{0}) \left( 
	\frac{M_{{\rm eq}}}{M_{0}} \right)^{(1-n)/6} \\ \nonumber
  & & \hspace*{1cm} \times \left( 
	\frac{M_{{\rm dec}}}{M_{{ \rm eq}}} \right)^{(1-n)/4} \,
	\left( \frac{M_{{\rm hor}}}{M_{{\rm dec}}} \right)^{(1-n)/6} \,,
\end{eqnarray}
where $M_0 \simeq 10^{56}\,$g is the present horizon mass. PBHs are formed
from rare, relatively large, density fluctuations which collapse soon after
entering the horizon, so we can take $M_{{\rm hor}} \sim M$.\footnote{A PBH 
forming well within the horizon will have a somewhat smaller mass than this 
(see e.g.~Ref.~\cite{CGL:PBH}), but when constraining the spectral index 
the correction is negligible.} This simplifies to 
\begin{equation}
\sigma_{{\rm hor}}(M)  =  \sigma_{{\rm hor}}(M_{0}) \left( \frac{M
         }{M_{0}}
	\right)^{(1-n)/6} \left( \frac{M_{{\rm dec}}}{M_{{\rm eq}}}
	\right)^{(1-n)/12} \,,
\end{equation}
and since during radiation domination $M_{{\rm H}} \propto T^{-2}$
\begin{eqnarray}
\label{sigd}
\sigma_{{\rm hor}}(M) & = & \sigma_{{\rm hor}}(M_{0}) \left( 
	\frac{M}{M_{0}} \right)^{(1-n)/6} \left( 
	\frac{T_{{\rm eq}}}{T_{{\rm dec}}} \right)^{(1-n)/6} \nonumber \\ 
  &  = & \sigma_{{\rm hor}}(M_{0}) \left(1.4\times10^{-6} 
  	\frac{M}{M_{0}} \right)^{(1-n)/6} \,,
\end{eqnarray}
for masses $M$ forming during moduli domination.

The lightest holes that can form are determined by the reheating temperature 
after the original period of inflation which is responsible for generating 
the density perturbations. The minimum mass is then given by 
Eq.~(\ref{mhor}). Normally, the tightest constraint on $n$ comes from the 
lightest PBHs. We use the method outlined in \cite{GL}, but using the 
expressions for $\sigma(M)$ and $\beta(M)$ given above, to obtain the 
constraints.

For immediate moduli domination, we find the tightest limit to be $n< 1.23$ 
from the deuterium constraint evaluated at $M \sim 10^{10}\,$g, although all 
the constraints due to the evaporation of PBHs require $n <1.32$. The limit 
from the present-day density of PBHs is tightest at $M \sim  5 \times 
10^{14}\,$g giving $n<1.30$. Relics do not constrain $n$, since even very 
large initial PBH abundances $\beta_{{\rm i}}$ will be diluted away. 

For our example case of delayed moduli domination, the most constraining
PBHs are formed during the radiation-dominated era before moduli domination
commences. For them $\sigma(M)$ has a different form
\begin{eqnarray}
\sigma_{{\rm hor}}(M) & =  & \sigma_{{\rm hor}}(M_{0}) \left( 
	\frac{M_{{\rm eq}}}{M_{0}} \right)^{(1-n)/6} \, \left( 
	\frac{M_{{\rm dec}}}{M_{{ \rm eq}}} \right)^{(1-n)/4} 
        \nonumber \\ 
 & & \hspace*{0.5cm} \times 
	\left( \frac{M_{\rm MD}}{M_{{\rm dec}}} \right)^{(1-n)/6} 
        \left( \frac{M}{M_{\rm MD}} \right)^{(1-n)/4} \,,
\end{eqnarray}
which simplifies to 
\begin{equation}
\sigma_{{\rm hor}}(M) = \sigma_{{\rm hor}}(M_{0}) \left(
	10^{5}\frac{M}{M_{0}} \right)^{(1-n)/4} \,,
\end{equation}
for our specific parameters. Since the PBHs of interest are formed during 
radiation domination the standard expression for $\beta$ applies 
\cite{Carr:rev}:
\begin{equation}
\beta(M) \approx \sigma(M) \exp{ \left( - \frac{1}{18 \sigma^2(M)}
	\right)} \,.
\end{equation}

\begin{figure}[t]
\centering 
\leavevmode\epsfysize=6cm \epsfbox{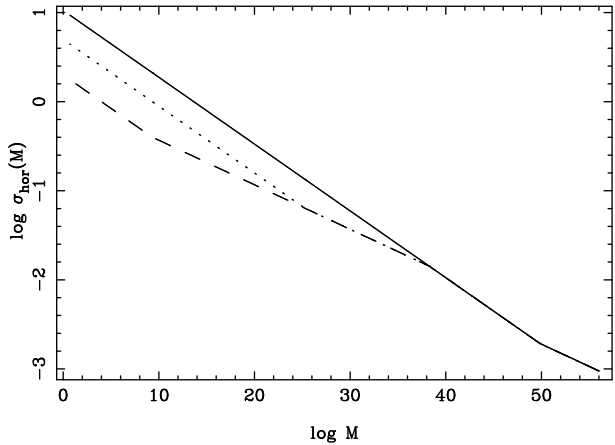}\\ 
\caption[fig4]{\label{fig4}  The variation of $\sigma_{{\rm hor}}(M)$
with mass for the three scenarios considered; the solid, dotted and
dashed lines representing the standard cosmology, immediate moduli
domination and delayed moduli domination respectively. 
The mass is in grams, and we take $n=1.3$ for illustrative purposes.}
\end{figure}

The tightest limit is now $n<1.26$ from the deuterium constraint evaluated
at $M \sim 10^{10}\,$g, with all the constraints due to the evaporation of 
PBHs require $n<1.28$. The tightest limit from the present-day density of
PBHs is $n<1.30$ at $M \sim 5 \times 10^{14}$g. The relic constraint may
provide an even tighter limit if the reheat temperature after inflation is
close to $10^{16}$ GeV.  

In Fig.~\ref{fig4} we illustrate the variation of $\sigma_{{\rm hor}}(M)$
in the standard cosmology, for immediate moduli domination and for our
example case of delayed moduli domination.

The tightest constraint for immediate moduli domination is only slightly
weaker than in the standard cosmology \cite{GL}, where the tightest
constraint (again deuterium) is $n<1.22$, with the evaporation constraints
all giving $n<1.24$. There, the relic constraint may give an even tighter
limit if the reheat temperature after inflation is high enough ($\gtrsim 
10^{14}\,$GeV) to let quite light PBHs form. The weakening is only small, 
since during matter domination PBHs form more readily so that to attain any 
particular value of $\beta_{{\rm i}}$ a smaller value of $\sigma(M)$, and 
hence $n$, is necessary. This reduces the effect of the larger value 
of $\beta_{{\rm i}}$ allowed due to the period of matter domination, and
also leads to a larger spread in the limits on $n$ from different sources. 
The tightest constraint is significantly weaker for delayed moduli 
domination, since in this case the most constraining PBHs are formed during 
radiation domination so that the main difference from the standard scenario 
is the larger values of $\beta_{{\rm i}}$ allowed.

\section{Conclusions}

If there is a prolonged period of matter-domination by moduli in the 
early Universe, it leads to a weakening of the constraint on density 
perturbations from primordial black hole formation. It again reminds us of 
the sensitivity of this bound to the entire assumed cosmological history. 
If the moduli dominate immediately, the fraction of the density of the 
Universe permitted to go into PBHs becomes of order $10^{-8}$, rather than 
the $10^{-20}$ or so which the standard cosmology requires. Delayed moduli 
domination leads to an intermediate constraint on those PBHs which form 
before moduli domination. This weakening is similar to that found 
\cite{GL} for the case where an extra period of inflation at low energies, 
known as thermal inflation, is assumed. 

When expressed as a limit on the spectral index of a power-law density 
perturbation spectrum, we obtain \mbox{$n \lesssim 1.3$} for immediate 
moduli domination, rather than $n \lesssim 1.25$ as in the standard 
cosmology. The weakening is similar to that from thermal inflation, which
also led to $n \lesssim 1.3$. Interestingly, the constraint can actually be 
weakest if moduli domination is delayed, because PBH formation is harder 
during radiation domination than moduli domination.

We end by noting that the assumption of gaussian perturbations in the black 
hole formation calculation has recently been questioned by Bullock and 
Primack \cite{BP:gauss}. As shown in Ref.~\cite{GL}, in the most 
non-gaussian case found by Bullock and Primack the constraint on $n$ can be 
weakened further, by up to 0.05.

\section*{Acknowledgments}

A.M.G.~was supported by PPARC, A.R.L.~by the Royal Society and A.R.~by the 
DOE and NASA under Grant NAG5--2788. We thank Beatriz de Carlos, Bernard 
Carr and Jim Lidsey for discussions, and A.M.G.~acknowledges use of the 
Starlink computer system at the University of Sussex.


\begin{references}

\bibitem{LS:TI} D. H. Lyth and E. D. Stewart, Phys. Rev. Lett. 
	{\bf 75}, 201 (1995).
\bibitem{moreTI} D. H. Lyth and E. D. Stewart, Phys. Rev. D {\bf
	53}, 1784 (1996).
\bibitem{var:lim} B. J. Carr, Astrophys. J. {\bf 205}, 1 (1975);
         Ya. B. Zel'dovich, A. A. Starobinsky, M. Y. Khlopov and V. M.
         Chechetkin, Pis'ma Astron. Zh. {\bf 3}, 308 (1977) [Sov
         Astron. Lett. {\bf 22}, 110 (1977)]; S. Mujana and K. Sato,
         Prog. Theor. Phys. {\bf 59},1012 (1978); B. V. Vainer and
         P. D. Nasselskii, Astron. Zh {\bf 55}, 231 (1978)
         [Sov. Astron. {\bf 22}, 138 (1978)]; B. V. Vainer, D. V
         Dryzhakova and P. D. Nasselskii, Pis'ma Astron. Zh. {\bf 4},
         344 (1978) [Sov. Astron. Lett {\bf 4},185 (1978)];
         I. D. Novikov, A. G. Polnarev A. A. Starobinsky and
         Ya. B. Zel'dovich, Astron. Astrophys. {\bf 80}, 104 (1979);
         D. Lindley, Mon. Not. R. Astron. Soc. {\bf 193}, 593 (1980);
         T. Rothman and R. Matzner, Astrophys. Space. Sci {\bf 75},
         229 (1981); J. H. MacGibbon and B. Carr, Astrophys. J. 
         {\bf 371}, 447 (1991).
\bibitem{Carr:rev} B. J. Carr, {\em Observational and Theoretical
	Aspects of Relativistic Astrophysics and Cosmology} edited by
	J. L. Sanz and L. J. Goicoechea (World Scientific, Singapore,
	1985).
\bibitem{GL} A. M. Green and A. R. Liddle, Sussex preprint 
	astro-ph/9704251.
\bibitem{sugra} For a review, see, H. P. Nilles, Phys. Rep. {\bf 110}, 1 
	(1984); H. E. Haber and G. L. Kane, Phys. Rep. {\bf 117}, 75 
	(1985); A. Chamseddine, R. Arnowitt and P. Nath, 
	{\it Applied N=1 Supergravity}, World Scientific, 
	Singapore (1984).
\bibitem{moduli1} B. de Carlos, J. A. Casas, F. Quevedo and E. Roulet, 
	Phys. Lett. B {\bf 318}, 447 (1993).
\bibitem{moduli2} T. Banks, D. Kaplan, and A. Nelson, Phys. Rev. D
	{\bf D49}, 779 (1994). 
\bibitem{attempts} For a review of such attempts, see T. Banks, M. 
	Berkooz and P. J. Steinhardt, Phys. Rev. D {\bf 52}, 705 (1995).
\bibitem{gb} P. Binetruy, M. K. Gaillard and Y. Wu, LBNL-39744 
	preprint, hep-th/9702105.
\bibitem{b} For a general discussion and references, see T. Banks and 
	M. Dine, SCIP-96-31 preprint, hep-ph/9608197. 
\bibitem{dis} A. S. Goncharev, A. D. Linde and M. I. Vysotsky, Phys. 
	Lett. {\bf B147}, 279 (1984).
\bibitem{sh} M. Dine, W. Fischler and D. Nemechansky, Phys. Lett. 
	{\bf B136}, 169 (1984); G. D. Coughlan, R. Holman, P. Ramond
	adn G. G. Ross, Phys. Lett. {\bf B140}, 44 (1984). 
\bibitem{Page:tevap} D. N. Page, Phys. Rev. D {\bf 13}, 198 (1976).
\bibitem{MacG} J. H. MacGibbon, Nature {\bf 320}, 308 (1987); J. D. 
	Barrow, E. J. Copeland and A. R. Liddle, Phys. Rev. D {\bf 46}, 
	645 (1992).
\bibitem{CGL:PBH} B. J. Carr, J. H. Gilbert and J. E. Lidsey,
	Phys. Rev. D {\bf 50}, 4853 (1994).
\bibitem{kim} H. I. Kim and C. H. Lee, Phys. Rev. D {\bf 54}, 6001 (1996).
\bibitem{cohere} B. Ratra, Phys. Rev. D {\bf 44}, 352 (1991); J. Hwang,
	Kyungpook preprint astro-ph/9610042.
\bibitem{YP:dust} M. Y. Khlopov and A. G. Polnarev, Phys. Lett {\bf 97B},
	383 (1980); A. G. Polnarev and M. Y. Khlopov, Sov. Astron 
	{\bf 26}, 391 (1983); Sov. Phys. Usp. {\bf 28}, 213 (1985).
\bibitem{BP:gauss} J. S. Bullock and J. R. Primack, to appear, Phys. Rev.
	D, astro-ph/9611106.
\end{references}
\end{document}